\documentclass{taira}

\usepackage{amssymb}
\usepackage{amsmath}
\usepackage{graphicx}
\usepackage{lipsum}
\usepackage{graphicx}
\usepackage{epstopdf, epsfig}
\usepackage{amsmath, amssymb}
\usepackage{multirow}
\usepackage{bm}
\usepackage{here}
\usepackage{graphicx,siunitx}
\usepackage{graphicx}
\usepackage{listings,color,courier,natbib}
\usepackage{pdflscape, csquotes}
\usepackage[abs]{overpic}

\usepackage[T1]{fontenc}
\usepackage{adjustbox}
\usepackage{rotating}

\shorttitle{Machine learning based spatio-temporal super resolution for turbulence}
\shortauthor{K. Fukami, K. Fukagata and K. Taira}

\title{Machine learning based spatio-temporal super resolution reconstruction of turbulent flows}

\author{Kai Fukami\aff{1,2}
  \corresp{\email{kfukami1@g.ucla.edu}},
  Koji Fukagata\aff{2}
 \and Kunihiko Taira\aff{1}}

\affiliation{
\aff{1}Department of Mechanical and Aerospace Engineering, University of California, Los Angeles, CA 90095, USA
\aff{2}Department of Mechanical Engineering, Keio University, Yokohama, 223-8522, Japan
}

\begin{document}

\maketitle

\begin{abstract}
We present a new data reconstruction method with supervised machine learning techniques inspired by $\it super~resolution$ and $\it inbetweening$ to recover high-resolution turbulent flows from grossly coarse flow data in space and time. 
For the present machine learning based data reconstruction, we use the downsampled skip-connection/multi-scale model based on a convolutional neural network, incorporating the multi-scale nature of fluid flows into its network structure. 
As an initial example, the model is applied to the two-dimensional cylinder wake at $Re_D$ = 100. 
The reconstructed flow fields by the present method show great agreement with the reference data obtained by direct numerical simulation. 
Next, we apply the current model to a two-dimensional decaying homogeneous isotropic turbulence. 
The machine-learned model is able to track the decaying evolution from spatial and temporal coarse input data. 
The proposed concept is further applied to a complex turbulent channel flow over a three-dimensional domain at $Re_{\tau}=180$.
The present model reconstructs high-resolved turbulent flows from very coarse input data in space, and also reproduces the temporal evolution for appropriately chosen time interval. 
The dependence on the amount of training snapshots and duration between the first and last frames based on a temporal two-point correlation coefficient are also assessed to reveal the capability and robustness of spatio-temporal super resolution reconstruction. 
These results suggest that the present method can perform a range of flow reconstructions in support of computational and experimental efforts.

\end{abstract}


\section{Introduction}

In recent years, machine learning methods have been utilized to tackle various problems in fluid dynamics \citep{BNK2019,FFT2020,BEF2019,BHT2020}.  
Applications of machine learning for turbulence modeling have been particularly active in fluid dynamics \citep{DIX2019,Kutz2017}.
\citet{LKT2016} proposed {a} tensor-basis neural network based on {the} multi-layer perceptron \citep[MLP;][]{RHW1986} for Reynolds-Averaged Navier--Stokes simulation.
Embed{ding the} Galilean invariance into {the} machine learning structure was found to be important and was verified by considering {their model for flows in a duct and over a wavy-wall}.  
For large eddy simulation (LES), subgrid modeling assisted by machine learning was proposed by \citet{MSRB2019}.  
They showed the capability of machine learning {assisted subgrid} model{ing} {in} {\it a priori} and {\it a posteriori} {tests} for the Kraichnan turbulence.

Furthermore, machine learning {is proving itself as} a promising tool for {developing} reduced order models {(ROMs)}.
{For instance, \citet{MFF2019} proposed a n}onlinear mode decomposition technique {using an autoencoder \citep{HS2006} based on} convolutional neural network{s} \citep[CNN;][]{LBBH1998} and demonstrated its use on transient and asymptotic laminar cylinder wakes at $Re_D=100$.
Their method shows potential of autoencoder in terms of the feature extraction of flow fields in lower dimension.
{More recently, {\citet{HFMF2019} combined a CNN} and {the} long short term memory \citep[LSTM;][]{HS1997}} for {developing an ROM for} a two-dimensional unsteady wake behind various bluff bodies.
Although the aforementioned examples {deal} with only laminar flows, the strengths of machine learning have been capitalized for ROM of turbulent flows.
\citet{SM2018} utilized an extreme learning machine \citep{HZS2004} based on MLP for developing {an ROM of} geophysical turbulence.
{\citet{SGASV2019} used LSTM to predict temporal evolution of} the coefficients of nine-equation {turbulent} shear flow model.
{They demonstrated that the chaotic behavior of those coefficients can be reproduced well.}
{They also confirmed that the} statistics collected from machine learning {agreed with the} reference data.

Of particular interest here for fluid dynamics is the use of machine learning as a powerful approximator \citep{Kreinovich1991,Hornik1991,Cybenko1989,BFK2018}{,} which can handle nonlinearities.
We recently proposed {a} super resolution reconstruction {method} for fluid flows, which was tested for two-dimensional laminar cylinder wake and two-dimensional decaying homogeneous isotropic turbulence \citep{FFT2019b}.  
We demonstrated that high resolution {two-dimensional} turbulent flow fields of $128\times 128$ grid can be reconstructed from the input data on a coarse $4\times 4$ grids via machine learning method. 
Applications and extensions of super resolution reconstruction can be considered for not only computational \citep{OSM2019,LTHL2020} but also experimental fluid dynamics \citep{DHLK2019,MFF2020}.
Although these {attempts} showed great potential {of machine learning based super resolution methods} to {handle} high-resolved fluid big data efficiently, {their applicability has been so far} limited only to two-dimensional {\it spatial} reconstruction.

{In addition, the temporal data interpolation with various data-driven techniques has been used to process video images, including the optical flow based interpolation \citep{ilg2017flownet}, phase-based interpolation \citep{meyer2018phasenet}, pixels motion transformation \citep{jiang2018super}, and neural network-based interpolation \citep{xu2020stochastic}.}
{For instance, \citet{LRT2019} developed} {a machine learning based temporal super resolution technique called }{\it inbetweening} to estimate the snapshot sequences {between} the start and last frames for image and video processing.  
They {generated 14 possible frames} between {two} frames of videos using machine learning to save on storage.
In the fluid dynamics community, a similar concept {has} recently {been} considered {by \citet{VKWHL2020}} to temporal data interpolation for PIV measurement.
They developed {a} model based on {the} rapid distortion theory and Taylor's hypothesis.

{In the present study}, we perform a machine learning based spatio-temporal super resolution analysis inspired by {the aforementioned spatial} {\it super resolution} and {temporal} {\it inbetweening} techniques to reconstruct high-resolution turbulent flow data from extremely low resolution flow data {both} in space and time.
The present paper is organized as follows. 
{We first introduce our} machine learning based spatio-temporal super resolution approach {in section 2} with a simple demonstration for a two-dimensional laminar cylinder wake at $Re_D=100$.  
We then apply the present method to two-dimensional decaying isotropic turbulence and turbulent channel flow over three-dimensional domain {in section 3}.
The capability of machine learning based spatio-temporal super resolution method is assessed statistically.
Finally, concluding remarks are provided in section 4. 

\section{Approach}
\subsection{Spatio-temporal super resolution flow reconstruction with machine learning}

\begin{figure}
	\begin{center}
		\includegraphics[width=1\textwidth]{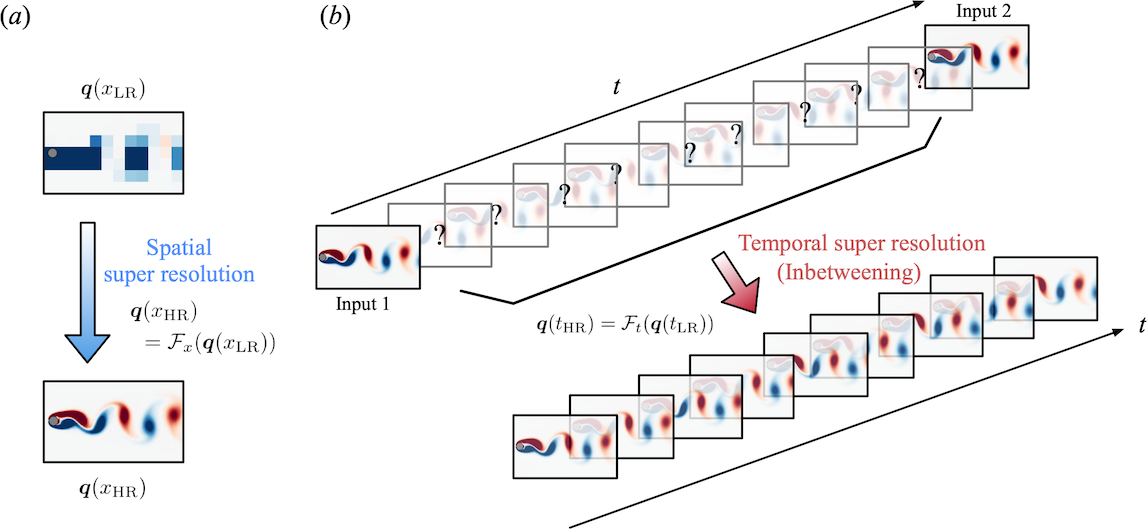}
		\caption{Data reconstruction methods used in the present study: $(a)$ Spatial super resolution. $(b)$ Temporal super resolution (inbetweening).}
		\label{fig1_20191122}
	\end{center}
\end{figure}

The objective of this work is to reconstruct high-resolution flow field data ${\bm q}(x_{\rm HR}, t_{\rm HR})$ from low-resolution data in space and time ${\bm q}(x_{\rm LR}, t_{\rm LR})$.  
To achieve this goal, we combine spatial {\it super resolution} analysis with temporal {\it inbetweening}.  
Super resolution analysis can reconstruct a spatially high-resolution data from spatially low-resolution input data, as illustrated in figure \ref{fig1_20191122}$(a)$.  
Temporal inbetweening is able to find the temporal sequences between the first and the last frames in the time-series data, as shown in figure \ref{fig1_20191122}$(b)$.  
We describe the methodology to combine these two reconstruction methods in section 2.2.

In the present study, we use a supervised machine learning model to reconstruct fluid flow data in space and time.
For supervised machine learning, we prepare a set of input {$\bm x$} and output (answer) {$\bm y$} as the training data.  
We then train the supervised machine learning model with these training data such that a nonlinear mapping function ${\bm y}\approx{\cal F}({\bm x};{\bm w})$ can be built, where $\bm w$ holds weights within the machine learning model.  
The training process here can be mathematically regarded as an optimization problem to determine the weights $\bm{w}$ such that ${\bm w}={\rm argmin}_{\bm w}[{E}({\bm y},{\mathcal F}({\bm x};{\bm w}))]$, where $E$ is the loss (cost) function.

\begin{figure}
	\begin{center}
		\includegraphics[width=1\textwidth]{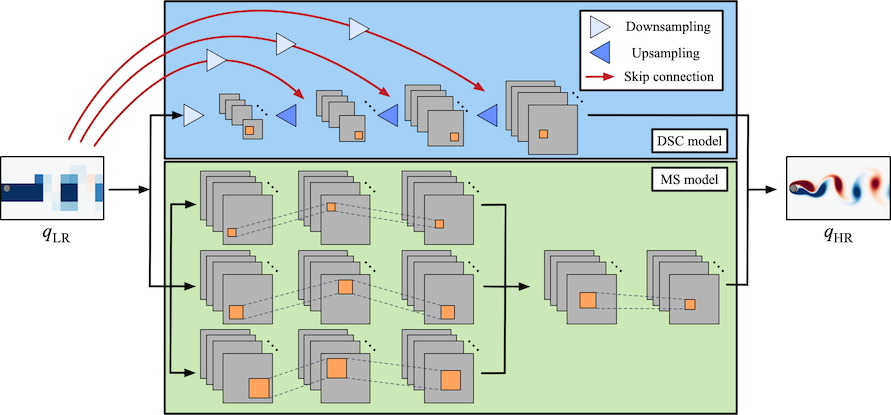}
		\caption{
		The hybrid down-sampled skip-connection/multi-scale (DSC/MS) super resolution model \citep{FFT2019b}.  Spatial reconstruction of two-dimensional cylinder wake at $Re_D=100$ is shown as an example.}
		\label{fig2_20191122}
	\end{center}
\end{figure}

For the machine learning models for super resolution in space ${\cal F}_x$ and time ${\cal F}_t$, we use a hybrid downsampled skip-connection and multi-scale (DSC/MS) model \citep{FFT2019b} presented in figure \ref{fig2_20191122}.
The DSC/MS model is based on a convolutional neural network \citep[CNN;][]{LBBH1998} which is one of the widely used supervised machine learning methods for image processing.
Here, let us briefly introduce the mathematical framework for the CNN.
The CNN is trained with filter operation such that
\begin{eqnarray}
    q^{(l)}_{ijm} = {\varphi}\biggl(b_m^{(l)}+\sum^{K-1}_{k=0}\sum^{H-1}_{p=0}\sum^{H-1}_{s=0}h_{p{s}km}^{(l)} q_{i+p-C,j+{s-C},k}^{(l-1)}\biggr),
\end{eqnarray}
where {$C={\rm floor}(H/2)$,} $b_m^{(l)}$ is the bias, $q^{(l)}$ is the output at layer $l$, $h$ is the filter, $K$ is the number of variables per each position of data, and $\varphi$ is an activation function which is generally chosen to be a monotonically increasing nonlinear function. 
In the present paper, we use the rectified linear unit (ReLU), $\varphi(s) = {\rm max}(0,s)$, as the activation function $\varphi$.
It is widely known that the use of ReLU enables machine learning models to be stable during the weight update process \citep{NH2010}.

As shown in figure \ref{fig2_20191122}, the present machine learning model is comprised of two models: namely the downsampled skip-connection model (DSC) model shown in blue and the multi-scale (MS) model shown in green.
The DSC model is robust against rotation and translation of the objects within the input images by combining compression procedures and skip-connection structures \citep{LNCCK2010,HZRS2016}.
On the other hand, the MS model \citep{DQHG2018} is able to take multi-scale properly of the flow field into account for its model structure.
Readers are refereed to \citet{FFT2019b} for additional details on the hybrid machine learning model.
The DSC/MS model is utilized for both spatial and temporal data reconstruction in the present study. 
{For the example of turbulent channel flow discussed in section \ref{sec:channel}, we use a three-dimensional convolution layer in place of the two-dimensional operations.}

\subsection{Order of spatio-temporal super resolution reconstruction}
\label{sec:order}

For the reconstruction of the flow field, we can consider the following two approaches:
\begin{enumerate}
    \item Apply the spatial super resolution model ${\cal F}_x^*:{\mathbb R}^{n_{\rm LR}\times m_{\rm LR}}\rightarrow {\mathbb R}^{n_{\rm HR}\times m_{\rm LR}}$, then the inbetweening model ${{\cal F}_t}:{\mathbb R}^{n_{\rm HR}\times m_{\rm LR}}\rightarrow {\mathbb R}^{n_{\rm HR}\times m_{\rm HR}}$ such that
    \begin{equation}
        {\bm q}(x_{\rm HR}, t_{\rm HR}) = {\cal F}_t({\cal F}_x^*({\bm q}(x_{\rm LR}, t_{\rm LR})))+{\bm \epsilon}_{tx}.
    \end{equation}
    \item Apply the inbetweening model ${\cal F}_t^*:{\mathbb R}^{n_{\rm LR}\times m_{\rm LR}}\rightarrow {\mathbb R}^{n_{\rm LR}\times m_{\rm HR}}$, then the spatial super resolution model ${{\cal F}_x}:{\mathbb R}^{n_{\rm LR}\times m_{\rm HR}}\rightarrow {\mathbb R}^{n_{\rm HR}\times m_{\rm HR}}$ such that
    \begin{equation}
        {\bm q}(x_{\rm HR}, t_{\rm HR}) = {\cal F}_x({\cal F}_t^*({\bm q}(x_{\rm LR}, t_{\rm LR})))+{\bm \epsilon}_{xt},
    \end{equation}
\end{enumerate}
where $n$ is a spatial dimension of data, $m$ is a temporal dimension of data, ${\bm \epsilon}_{tx}$ is the error for the first case, and ${\bm \epsilon}_{xt}$ is the error for the second case.
The subscripts LR and HR denote low-resolution and high-resolution variables, respectively.

\begin{figure}
	\begin{center}
		\includegraphics[width=1\textwidth]{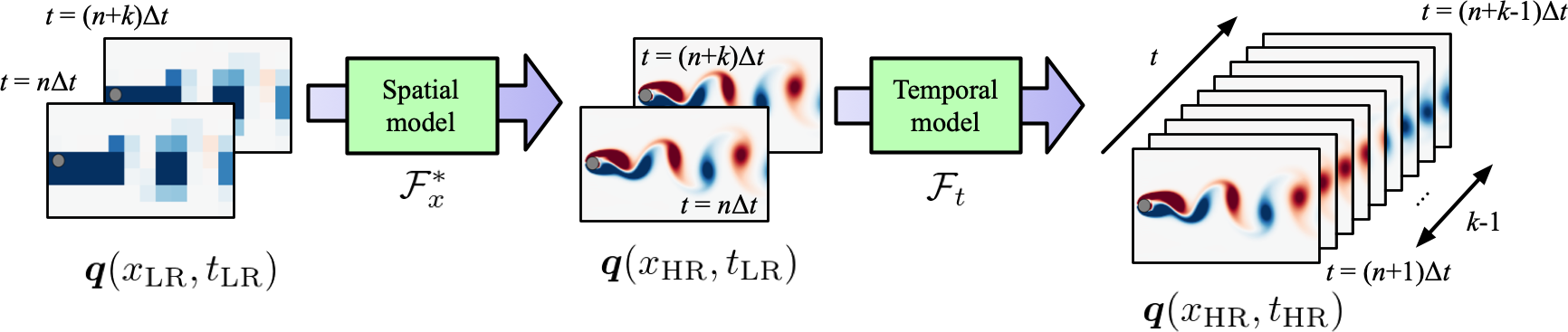}
		\caption{
		Spatio-temporal super resolution reconstruction with machine learning for cylinder flow at $Re_D = 100$.}
		\label{fig1}
	\end{center}
\end{figure}

\begin{figure}
	\begin{center}
		\includegraphics[width=0.70\textwidth]{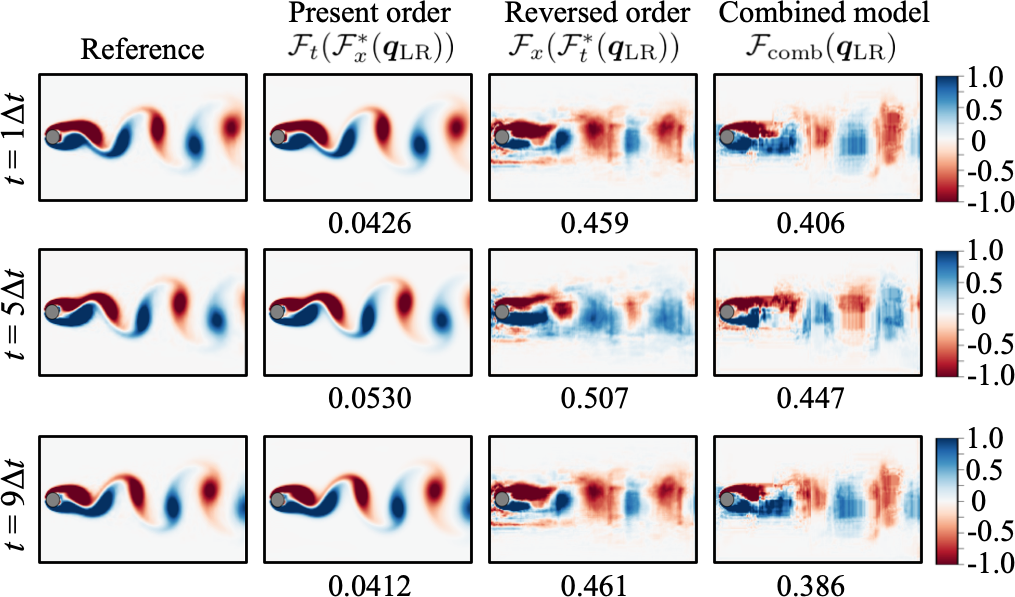}
		\caption{{
		Dependence of reconstruction performance on the order of training processes for cylinder flow. The first ($t=1\Delta t$), intermediate ($t=5\Delta t$), and last ($t=9\Delta t$) snapshots are shown. Reversed order refers to option (ii) in section \ref{sec:order}. The combined model refers to ${\cal F}_{\rm comb}$ which directly reconstructs ${\bm q}(x_{\rm HR}, t_{\rm HR})$ from ${\bm q}(x_{\rm LR}, t_{\rm LR})$. The values underneath the contours report the $L_2$ error norms.
		}}
		\label{fig4_20200720}
	\end{center}
\end{figure}

We seek the approach that achieves lower error between the above two formulations.
The $L_p$ norms of these error are assessed as
\begin{eqnarray}
    ||{\bm \epsilon}_{tx}||_p 
    &=& ||{\bm q}(x_{\rm HR},t_{\rm HR})-{\cal F}_t({\cal F}_x^*({\bm q}(x_{\rm LR},t_{\rm LR})))||_p \nonumber\\
    &=& ||{\bm q}(x_{\rm HR},t_{\rm HR})-{\cal F}_t({\bm q}(x_{\rm HR},t_{\rm LR})+{\bm \epsilon}_x)||_p,\\
    ||{\bm \epsilon}_{xt}||_p
    &=& ||{\bm q}(x_{\rm HR},t_{\rm HR})-{\cal F}_x({\cal F}_t^*({\bm q}(x_{\rm LR},t_{\rm LR})))||_p \nonumber\\
    &=& ||{\bm q}(x_{\rm HR},t_{\rm HR})-{\cal F}_x({\bm q}(x_{\rm LR},t_{\rm HR})+{\bm \epsilon}_t)||_p,
\end{eqnarray}
where ${\bm \epsilon}_x$ is an error from the spatial super resolution algorithm for the first case and ${\bm \epsilon}_t$ is an error from the inbetweening process for the second case.  
Since spatial super resolution algorithm is not a function of the temporal resolution algorithm in our problem setting, ${\bm \epsilon}_x$ is not affected {much} by {the temporal coarseness of the given data}.
On the other hand, ${\bm \epsilon}_t$ is the error resulting from inbetweening with spatial low-resolution data which lacks the phase information compared to the spatially high-resolution data.  
For this reason, the error ${\bm \epsilon}_t$ is likely to be large due to the spatial coarseness.  
This leads us to first establish a machine learning model for spatio-temporal super resolution reconstruction as illustrated in figure \ref{fig1} for the example of a cylinder wake.  
{The cylinder flow example confirms the above error trend, as shown in figure \ref{fig4_20200720}}.
{We also examine the possibility of utilizing a single combined model $\cal F_{\rm comb}$, i.e., ${\bm q}(x_{\rm HR}, t_{\rm HR})={\cal F}_{\rm comb}({\bm q}(x_{\rm LR}, t_{\rm LR}))$, which attempts to reconstruct the spatio-temporal high-resolution flow field from its counterpart directly.
As presented in figure \ref{fig4_20200720}, the flow field cannot be reconstructed well.
This is caused by the difficulty in weight updates while training the machine learning model.
The observations here suggest that care should be taken in training machine learning models \citep{FNF2020}.
}

Each of these machine learning models is trained individually for the spatial and temporal super resolution reconstructions.  
In the supervised machine learning process for regression tasks, the training process is formulated as an optimization problem to minimize a loss function in an iterative manner.  
The objectives of the two machine learning models can be expressed as
\begin{eqnarray}
    {\bm w}_{x} 
    = {\rm argmin}_{{\bm w}_x}
    ||{\bm q}(x_{\rm HR},t_{\rm LR})-{\cal F}_x^*({\bm q}(x_{\rm LR},t_{\rm LR}))||_p,\\
    {\bm w}_{t} 
    = {\rm argmin}_{{\bm w}_t}
    ||{\bm q}(x_{\rm HR},t_{\rm HR})-{\cal F}_t({\bm q}(x_{\rm HR},t_{\rm LR}))||_p,
\end{eqnarray}
where ${\bm w}_{x}$ and ${\bm w}_{t}$ are the weights of the spatial and temporal super resolution models, respectively.  
In the present study, we use the $L_2$ norm (${p}=2$) to determine the optimized weights $\bm w$ for each of the machine learning models. 
Hereafter, we use ${p}=2$ for assessing the errors.

\subsection{Demonstration: two-dimensional laminar cylinder wake}

For demonstration, let us apply the proposed formulation to the two-dimensional cylinder wake at $Re_D=100$.  
The snapshots for this wake are generated by two-dimensional direct numerical simulation (DNS) \citep{TC2007,CT2008}, which numerically solves the incompressible Navier--Stokes equations,
\begin{eqnarray}
\nabla \cdot {\bm u}= 0, \\
\frac{\partial {\bm u}}{\partial t}+{\bm u} \cdot \nabla {\bm u}=-\nabla p+\frac{1}{Re_D}\nabla^2\bm u.
\end{eqnarray}
{Here $\bm u$ and $p$ are the non-dimensionalized velocity vector and pressure, respectively.  
All variables are made dimensionless by the fluid density $\rho$, the uniform velocity $U_\infty$, and the cylinder diameter $D$.
The Reynolds number is defined as $Re_D=U_\infty D/\nu$ with $\nu$ being the kinematic viscosity.}
For this example, we use five nested levels of multi-domains with the finest level being $(x, y)/D = [-1, 15] \times [-8, 18]$ and the largest domain being $(x,y)/D = [-5, 75] \times [-40, 40]$.  Each domain uses {$[N_x, N_y] = [400, 400]$ for discretization.}
The time step for DNS is set to $\Delta t=2.50\times 10^{-3}$ and yielding a maximum CFL number of 0.3. 
As the training data set, we extract the domain around a cylinder body over $(x^*,y^*)/D = [-0.7, 15] \times [-5, 5]$ with $(N_x,N_y) = (192, 112)$.  
For the present study, we use 70\% of the snapshots for training and the remaining 30\% for validation{, which splits the whole data set randomly.}
{The assessment of this demonstration is performed using 100 test snapshots excluding the training and validation data.
Note here that the nature of training and test data sets are similar to each other due to the periodicity of the laminar two-dimensional circular cylinder wake.}
An early stopping criterion \citep{prechelt1998} with 20 iterations of the learning process is also utilized to avoid overfitting such that the model retains generality for any unseen data in the training process. 
{For the input and output attributes to the machine learning model, we choose the vorticity field $\omega$.}

\begin{figure}
	\begin{center}
		\includegraphics[width=1\textwidth]{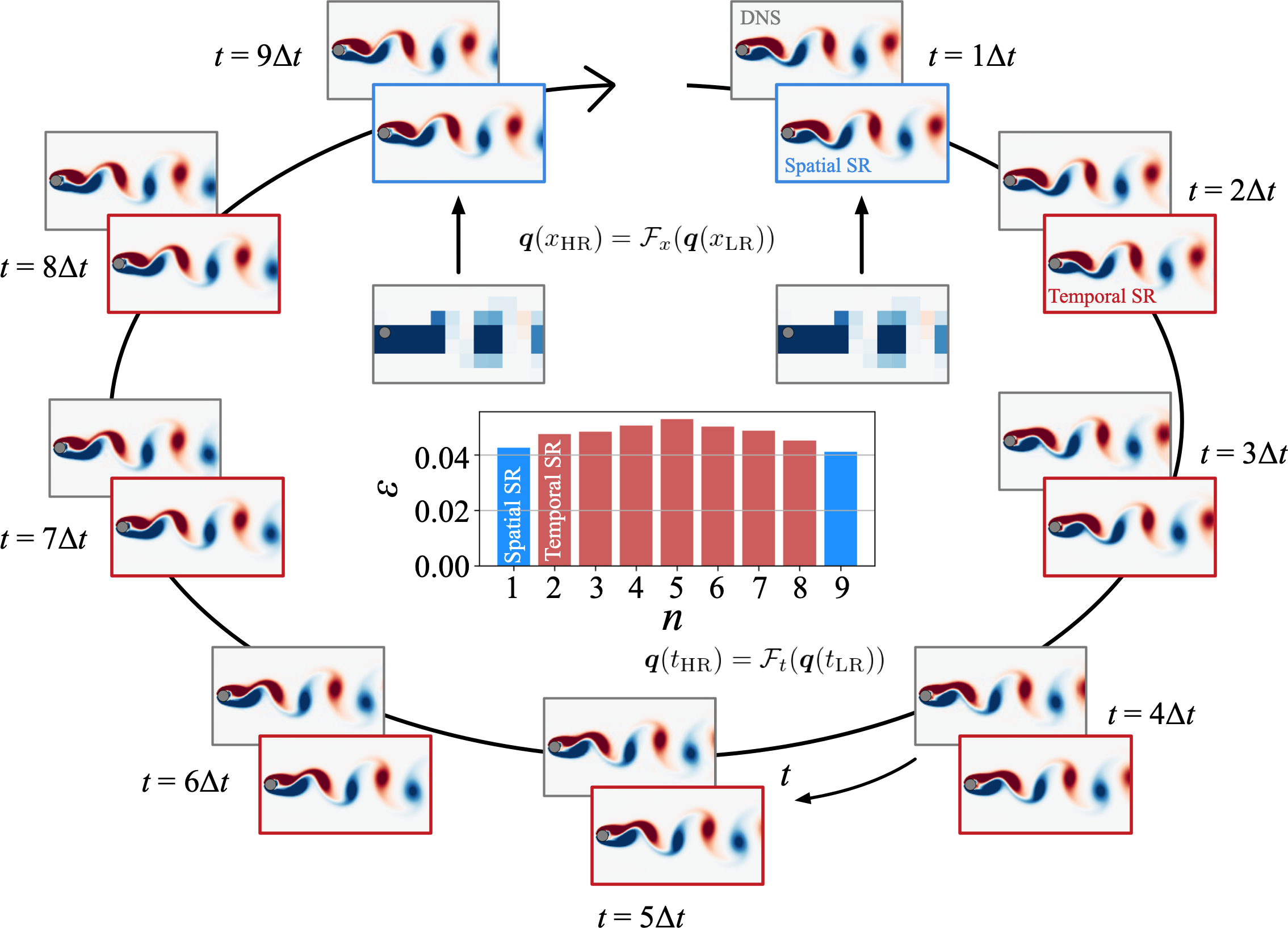}
		\caption{{
		Spatio-temporal super resolution reconstruction with machine learning for cylinder wake at $Re_D=100$.  The bar graph located on the center shows the $L_2$ error norm $\epsilon$ for the reconstructed flow fields. SR indicates super resolution. The contour level of the vorticity fields is same as that in figure \ref{fig4_20200720}.}}
		\label{fig2}
	\end{center}
\end{figure}

The results from preliminary examination with the undersampled cylinder wake data are summarized in figure \ref{fig2}. 
The machine learning models are trained by using $n_{{\rm snapshot}, x}=1000$ for spatial super resolution and $n_{{\rm snapshot}, t}=100$ for inbetweening.
{Both snapshots are prepared from a same time range which is approximately eight vortex shedding periods.}
Here, the spatial super resolution model ${\cal F}_x$ has the role of a mapping function from the low spatial resolution data ${\bm q}(x_{\rm LR})\in {\mathbb R}^{12\times 7}$ to the high spatial resolution data ${\bm q}(x_{\rm HR})\in {\mathbb R}^{192\times 112}$.
Next, two spatial high-resolved flow fields illustrated only at $t=1\Delta t$ and $9\Delta t$ in figure \ref{fig2} are used as the input for the temporal super resolution model ${\cal F}_t$, so that the inbetween snapshots from $t=2\Delta t$ to $8\Delta t$ corresponding to a period in time can be reconstructed.  
As shown in figure \ref{fig2}, the spatio-temporal super resolution analysis achieves excellent reconstruction of the flow field that is practically indistinguishable from the reference DNS data.  
The $L_2$ error norm {$\epsilon = ||{\omega}_{\rm DNS}-{\omega}_{\rm ML}||_2/||{\omega}_{\rm DNS}||_2$} is shown in the middle of figure \ref{fig2}.  
The $L_2$ error level is approximately 5\% of the reference DNS data.  
As the machine learning model is provided with the information at $t = 1\Delta t$ and $9 \Delta t$, the error level shows slight increase between those two snapshots.

\section{Results}

\subsection{Example 1: two-dimensional decaying homogeneous isotropic turbulence}

As the first example {of turbulent flows}, let us consider two-dimensional decaying homogeneous isotropic turbulence.  
The training data set is obtained by numerically solving the two-dimensional vorticity transport equation,
\begin{equation}
\frac{\partial \omega}{\partial t}+{\bm u}\cdot\nabla\omega=\frac{1}{Re_0}\nabla^2 \omega,
\label{eq_1}
\end{equation} 
where ${\bm u}=(u,v)$ and $\omega$ are the velocity and vorticity, respectively \citep{TNB2016}.  
The size of the biperiodic computational domain and the numbers of grid points here are $L_x=L_y=1$ and $N_x=N_y=128$, respectively. 
The Reynolds number is defined as $Re_0 \equiv u^*l_0^*/\nu$, where $u^*$ is the characteristic velocity obtained by the square root of the spatially averaged initial kinetic energy, {$l_0^*=[2{\overline{u^2}}(t_0)/{\overline{\omega^2}}(t_0)]^{1/2}$} is the initial integral length, and $\nu$ is the kinematic viscosity.  
The initial Reynolds numbers are $Re_0 = u^*(t_0)l^*(t_0)/\nu=$ 81.2 for training/validation data and 85.4 for test data {\citep{FFT2019b}}.  
For the input and output attributes to the machine learning model, we use the vorticity field $\omega$.

\begin{figure}
	\begin{center}
		\includegraphics[width=1\textwidth]{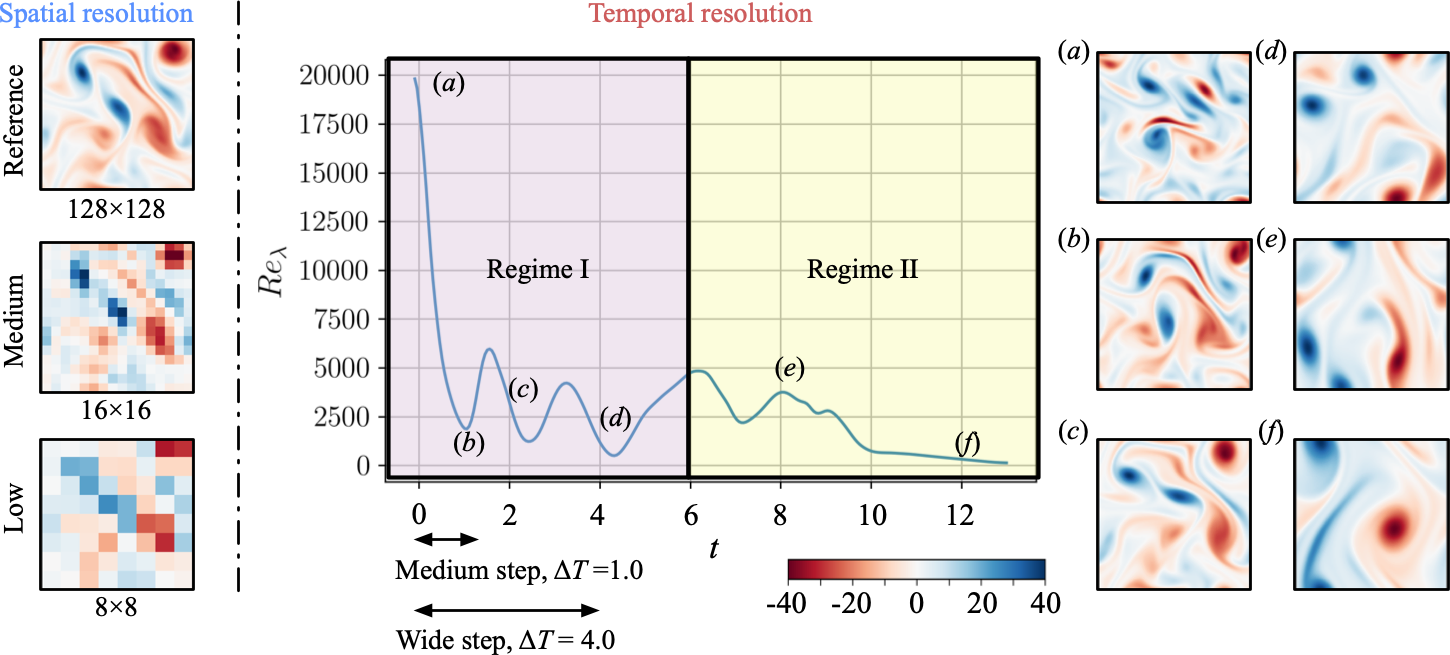}
		\caption{{
		The problem set up of spatio-temporal super resolution analysis for two-dimensional decaying homogeneous isotropic turbulence. The curve located on center is the decay of Taylor Reynolds number $Re_{\lambda}$. The vorticity fields $\omega$ are shown.}
		}
		\label{fig3_n}
	\end{center}
\end{figure}

For spatio-temporal super resolution analysis of two-dimensional turbulence, we consider four cases comprised of two spatial and two temporal coarseness levels as shown in figure \ref{fig3_n}.
For spatial super resolution analysis, we prepare two levels of spatial coarseness: medium- ($16\times 16$) and low-resolution ($8\times 8$ grids) data, analogous to our previous work \citep{FFT2019b}.  
These spatial low resolution data are obtained by an average downsampling of the reference DNS data set.  
{Note that the reconstruction with super resolution for turbulent flows is influenced by the downsampling process, e.g., max or average values \citep{FFT2019b}.
We use the average pooling in the present work.}
For the temporal resolution set up, we define medium- ($\Delta T=1.0$) and wide-time step ($\Delta T=4.0$), where $\Delta T$ is the time step between the first and last frames of the inbetweening analysis. 
{The decay of Taylor Reynolds number $Re_\lambda(t) = u^{\#}(t)\lambda(t)/\nu$, where $u^{\#}(t)$ is the spatial root-mean-square value for velocity at an instantaneous field and $\lambda(t)$ is the Taylor length scale at an instantaneous field, is shown in the middle of figure \ref{fig3_n}.}
The training data includes the low Taylor Reynolds number portion (Regime I\hspace{-.1em}I in figure \ref{fig3_n}) so as to assess the influence on the decaying physics compared to Regime I.
For the training process, we consider a fixed number of snapshots $(n_{{\rm snapshot}, x},n_{{\rm snapshot}, t})=(10000,10000)$ for {all four cases of} this two-dimensional example.
{The models for the considered four cases are constructed separately.}

\begin{figure}
	\begin{center}
		\includegraphics[width=1\textwidth]{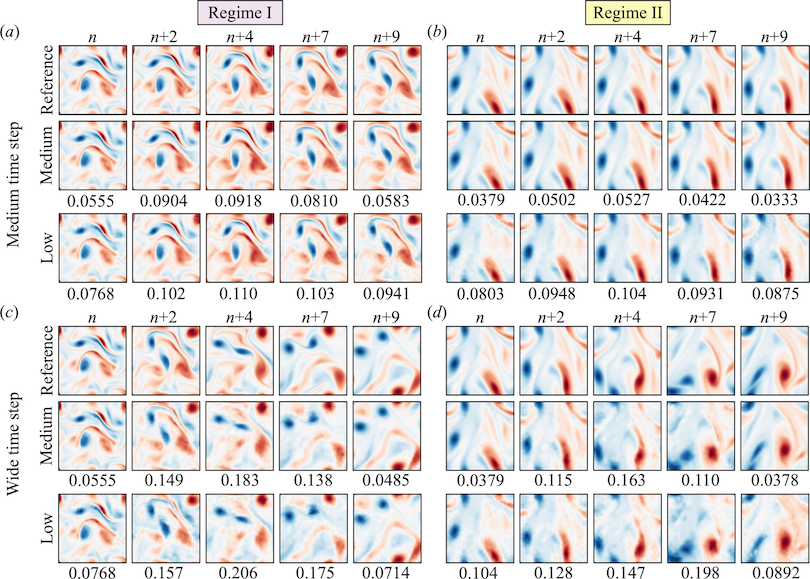}
		\caption{Spatio-temporal super resolution reconstruction for two-dimensional decaying homogeneous turbulence. {The vorticity field $\omega$ is shown with the same contour level in figure \ref{fig3_n}.} Medium- and low-resolution spatial input with $(a)$ medium time step for Regime I, $(b)$ medium time step for Regime I\hspace{-.1em}I, $(c)$ wide time step for Regime I, and $(d)$ wide time step for Regime I\hspace{-.1em}I.  The values underneath the flow fields report the $L_2$ error norms.}
		\label{fig4_n}
	\end{center}
\end{figure}

For the example of two-dimensional turbulence, the machine learning model for inbetweening analysis plays the role of a regression function to reconstruct 8 snapshots between the first and last frames (given by the spatial reconstruction model).  
The flow fields reconstructed from spatio-temporal super resolution analysis of two-dimensional turbulence with various coarse input data are summarized in figure \ref{fig4_n}.  
On the left side, the reconstructed fields from Regime I with coarse spatio-temporal data are shown.
As it can be seen, the temporal evolution of the complex vortex dynamics can be accurately reconstructed by the machine-learned models.
For almost all cases, the $L_2$ error norms $\epsilon = ||{\bm q}_{\rm DNS}-{\bm q}_{\rm ML}||_2/||{\bm q}_{\rm DNS}||_2$ listed below the contour plots for the spatially low-resolution input show larger errors compared to the medium-resolution case due to the effect of input spatial coarseness.
On the right side of the figure, we show the results from Regime I\hspace{-.1em}I.  
Analogous to the results for Regime I, the reconstructed flow fields are in agreement with the reference DNS data.  
Noteworthy here is the peak $L_2$ error norm of 0.198 appearing at $t=(n+7)\Delta t$ for Regime I\hspace{-.1em}I using low resolution input and a wide time step.
This is in contrast with the other cases that give peak errors at $t=(n+4)\Delta t$.  
{This is likely because the present machine learning model, which does not embed information of boundary condition, has to handle the temporal evolution {of a relatively large structure} over a bi-periodic domain (i.e., bottom left on the color map).}
Furthermore, the model is also affected by the error from the spatial super resolution reconstruction.
For these reasons, the peak in error is shifted in time compared to the other cases in figure \ref{fig4_n}.

\begin{figure}
	\begin{center}
		\includegraphics[width=0.7\textwidth]{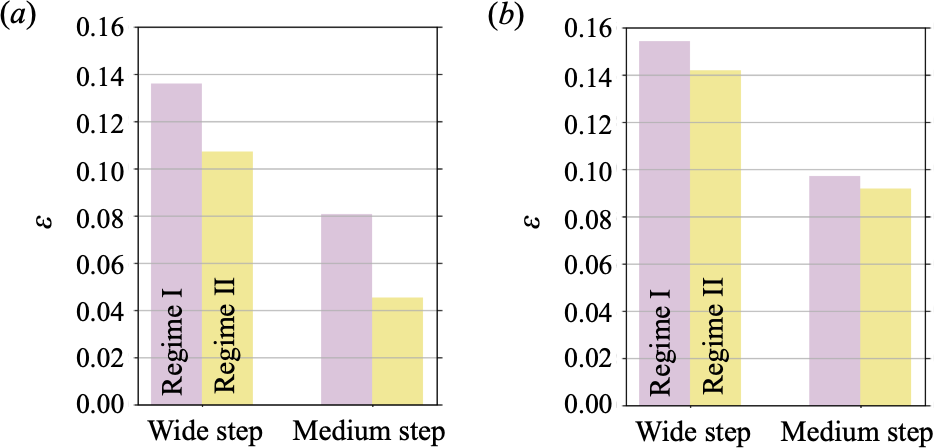}
		\caption{Time-ensemble $L_2$ error norms for the $(a)$ medium- and $(b)$ low-spatial input cases.}
		\label{fig5_n}
	\end{center}
\end{figure}

\begin{figure}
	\begin{center}
		\includegraphics[width=0.8\textwidth]{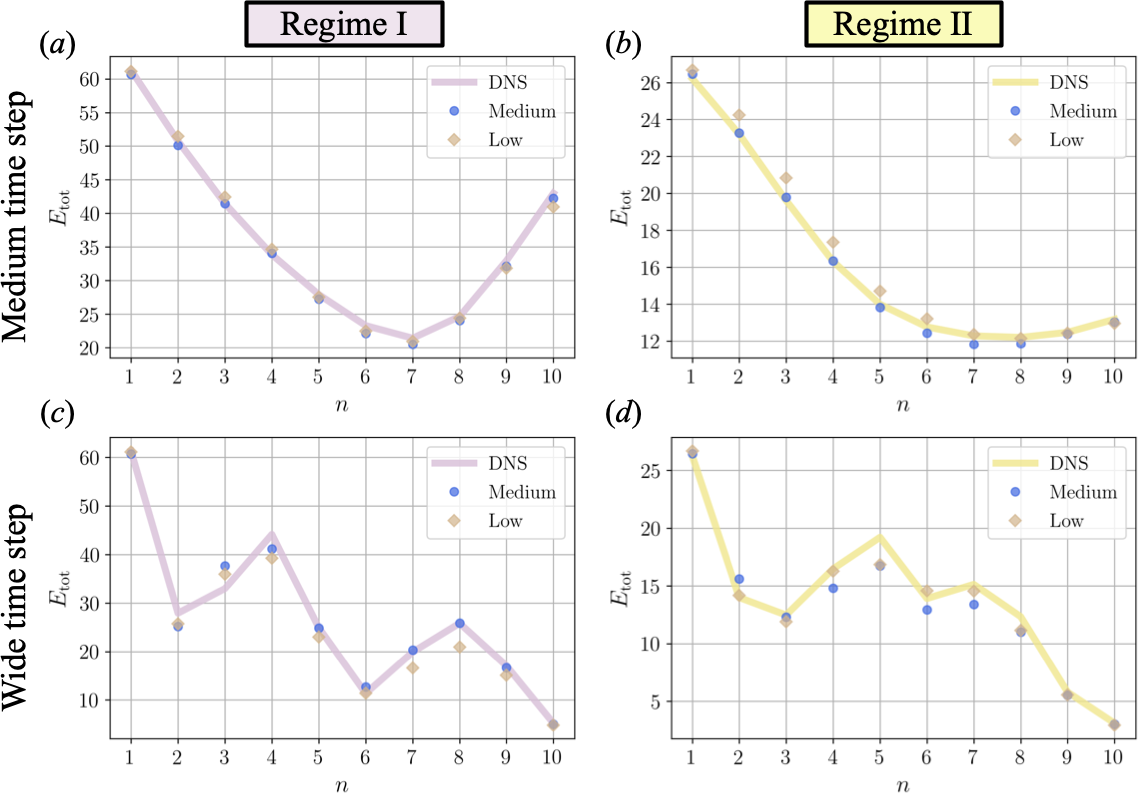}
		\caption{{Decay of time-ensemble total kinetic energy $E_{\rm tot}$ over the domain.}}
		\label{fig9_20200721}
	\end{center}
\end{figure}

To examine the dependence on the regime of test data, the time-ensemble $L_2$ error norms of medium- and low spatial input cases are shown in figure \ref{fig5_n}.  
For all cases, the errors for Regime I are larger than those for Regime I\hspace{-.1em}I.
One of reasons here is that the relative change in vortex structures for Regime I\hspace{-.1em}I is less than that for Regime I, which we can see in figure \ref{fig4_n}.
We also find that the reconstructions are affected by the input coarseness in space as evident from comparing figures \ref{fig5_n}$(a)$ and $(b)$.
{Similar trends can also be seen in figure \ref{fig9_20200721} which shows the total kinetic energy $E_{\rm tot}$ over the domain for each case.
The curves shown here do not monotonically decrease since we take the ensemble average over the test data, whose trend is analogous to the decay of Taylor Reynolds number as presented in figure \ref{fig3_n}.
By comparing figures \ref{fig4_n} and \ref{fig9_20200721}, it is inferred that the present machine learning model can capture the decaying nature of two-dimensional turbulence.}

\begin{figure}
	\begin{center}
		\includegraphics[width=1\textwidth]{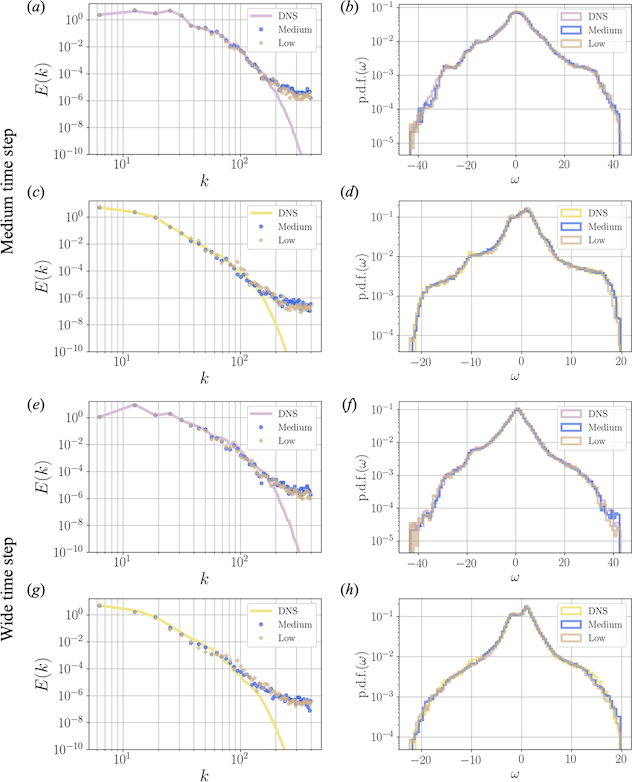}
		\caption{Statistical assessments for the spatio-temporal super resolution analysis of two-dimensional turbulence.  $(a),(c),(e)$, and $(g)$: Kinetic energy spectrum. $(b),(d),(f)$, and $(h)$: Probability density function of vorticity $\omega$. $(a)$--$(d)$: Medium time step.  $(e)$--$(f)$: Wide time step. $(a),(b),(e)$, and $(f)$: Regime I. $(c),(d),(g)$, and $(h)$: Regime I\hspace{-.1em}I.}
		\label{fig66_n}
	\end{center}
\end{figure}

Next, let us present the kinetic energy spectrum and the probability density function of the vorticity field $\omega$ for all coarse input cases with spatio-temporal reconstructions in figure \ref{fig66_n}.  
For comparison, we compute these statistics for Regimes I (purple) and I\hspace{-.1em}I (yellow).
The statistics with all coarse input data show similar distributions with the reference DNS trends.
The high wavenumber region of the kinetic energy spectrum obtained from the reconstructed fields do not match with the reference curve due to the lack of correlation between the low and high wavenumber components.
Through our investigation in this section, we can exemplify that the present machine learning model can successfully work in reconstructing high-resolution two-dimensional turbulence from spatio-temporal low-resolution data.

\subsection{Example 2: turbulent channel flow over three-dimensional domain}
\label{sec:channel}

\begin{figure}
	\begin{center}
		\includegraphics[width=1\textwidth]{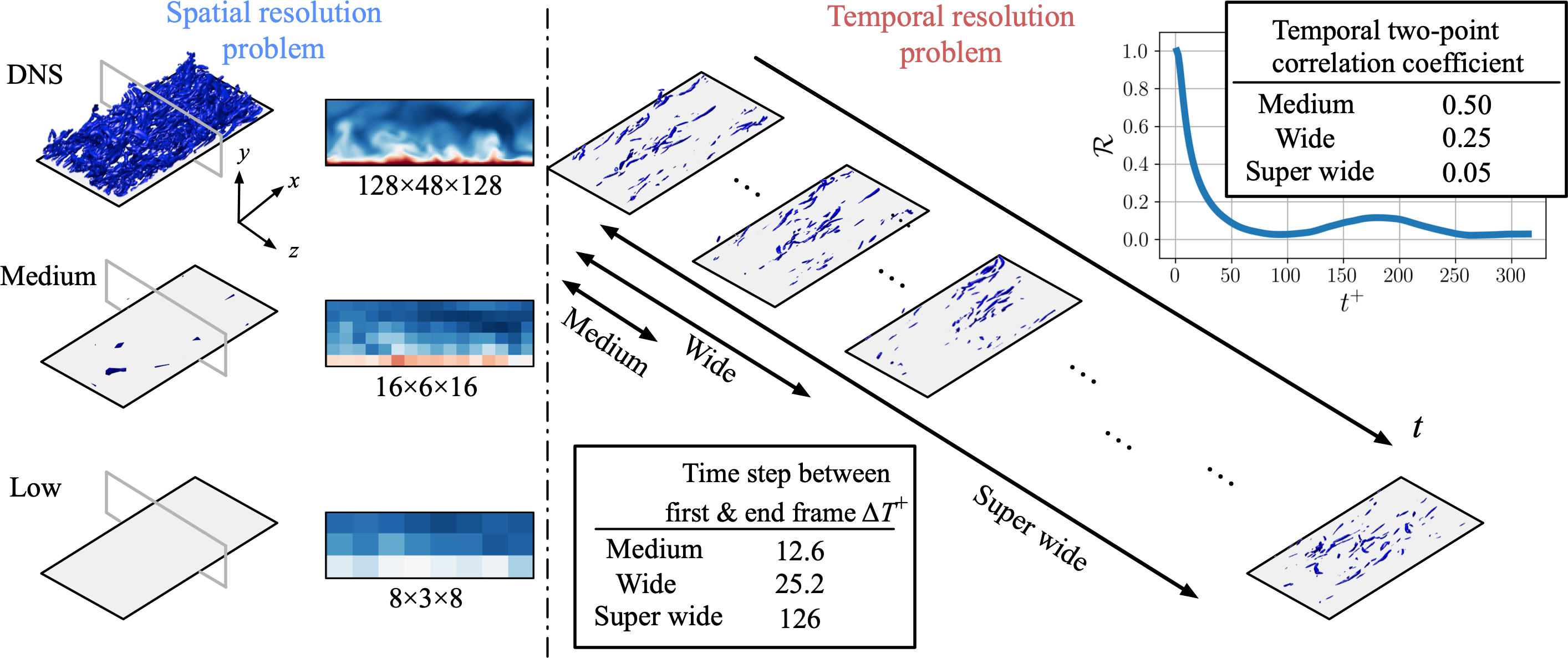}
		\caption{The problem set up for example 2.  We consider two spatial coarseness levels with three temporal resolutions.  Note that $Q^+=0.005$ and 0.07 are used for visualization of spatial and temporal resolutions, respectively.  The plot on the upper right shows the temporal two-point correlation coefficients $\cal R$ at $y^+=11.8$ for the present turbulent channel flow.}
		\label{fig6_n}
	\end{center}
\end{figure}

To investigate the applicability of machine learning based spatio-temporal super resolution reconstruction to three-dimensional turbulence, let us consider a turbulent channel flow \citep{FKK2006}.  
The governing equations are the incompressible Navier--Stokes equations,
\begin{eqnarray} 
\bm{\nabla} \cdot {\bm u} = 0,\\
{ \dfrac{\partial {\bm u}}{\partial t}  + \bm{\nabla} \cdot ({\bm u \bm u}) =  -\bm{\nabla} p  + \dfrac{1}{{Re}_\tau}\nabla^2 {\bm u}},
\end{eqnarray}
where $\displaystyle{{\bm u} = [u~v~w]^{\mathrm T}}$ represents the velocity vector with components $u$, $v$ and $w$ in the streamwise ($x$), wall-normal ($y$) and spanwise ($z$) directions.  
Here, $t$ is time, $p$ is pressure, and $\displaystyle{{Re}_\tau = u_\tau  \delta/\nu}$ is the friction Reynolds number.  
The variables are normalized by the half-width $\delta$ of the channel and the friction velocity {$u_\tau=(\nu {dU}/{dy}|_{y=0})^{1/2}$, where $U$ is the mean velocity}.  
The size of the computational domain and the number of grid points here are $(L_{x}, L_{y}, L_{z}) = (4\pi\delta, 2\delta, 2\pi\delta)$ and $(N_{x}, N_{y}, N_{z}) = (256, 96, 256)$, respectively.  
The grids in the $x$ and $z$ directions are taken to be uniform.
A non-uniform grid is utilized in the $y$ direction { with stretching based on the hyperbolic tangent function.}

As the baseline data, we prepare the data snapshots on a uniform grid interpolated from the non-uniform grid data of DNS.
The influence of grid type is reported in Appendix for completeness.  
The no-slip boundary condition is imposed on the walls and a periodic boundary condition is prescribed in the $x$ and $z$ directions.
The flow is driven by a constant pressure gradient at ${Re}_{\tau}=180$.  
For the present study, a subspace of the whole computational domain is extracted and used for the training process, i.e., $(L_x^*,L_y^*,L_z^*)=(2\pi\delta,\delta,\pi\delta)$, $x, y, z \in [0,L_x^*] \times [0,L_y^*] \times [0,L_z^*]$, and $(N_{x}^*, N_{y}^*, N_{z}^*) = (128, 48, 128)$.  
Due to the symmetry of turbulence statistics in the $y$ direction and homogeneity in the $x$ and $z$ directions, the extracted subdomain maintains the turbulent characteristics of the channel flow over the original domain size.
We generally use 100 training data set for both the spatial and temporal super resolution analyses in this case.
The dependence of the reconstruction on the number of snapshots is investigated later.
{For all assessments in this example, we use 200 test snapshots excluding the training data.}
For the input and output attributes to the machine learning model, we use the velocity fields $\displaystyle{{\bm u} = [u~v~w]^{\mathrm T}}$.
{Hereafter, superscript $+$ is used to denote quantities in wall units.}

\begin{figure}
		\begin{center}
			\includegraphics[width=1.0\textwidth]{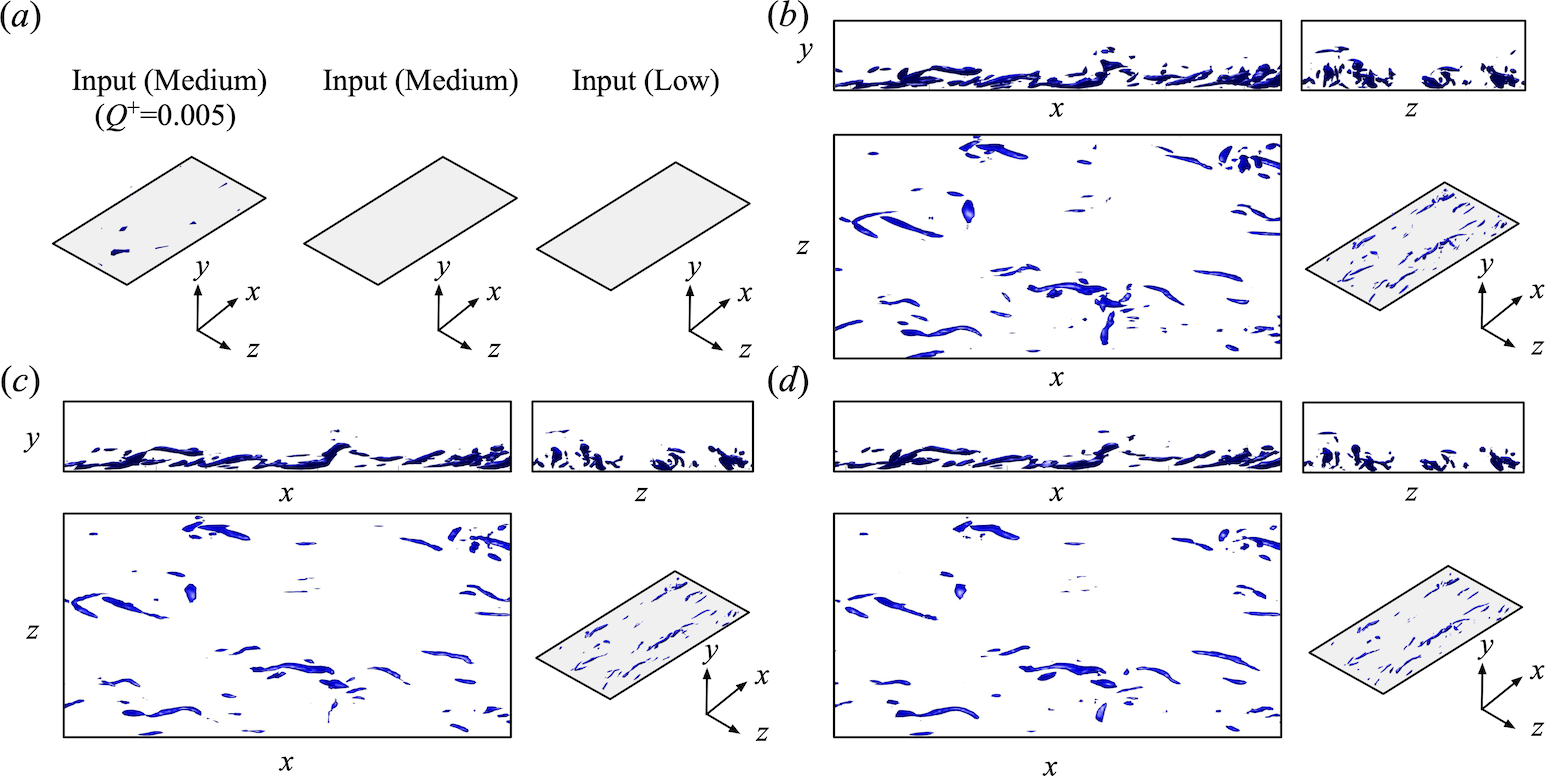}
			\caption{Isosurfaces of the $Q$ criterion ($Q^+=0.07$).  $(a)$ The input coarse data with medium and low resolutions.  For comparison, $Q^+=0.005$ with medium resolution is also shown.  $(b)$ Reference DNS data.  $(c)$ Reconstructed flow field from medium resolution input data.  $(d)$ Reconstructed flow field from low resolution input data.}
			\label{fig3}
		\end{center}
	\end{figure}

We illustrate in figure \ref{fig6_n} the problem setting of the spatio-temporal super resolution analysis for three-dimensional turbulence.
{For visualization here, we use the second invariant of the velocity gradient tensor $Q$.}
Regarding the spatial resolution, medium and low resolutions are defined as $16\times6\times16$ and $8\times3\times8$ grids in the $x$, $y$, and $z$ directions, respectively.  
These coarse input data sets are generated by the average downsampling operation from the reference DNS data of $128\times48\times128$ grids.  
Note that we are unable to detect the vortex core structures at $Q^+=0.005$ with low-resolution input in figure \ref{fig6_n} due to the gross coarseness.  
As shown in figure \ref{fig3}, the vortex structure cannot be seen with a contour level of $Q^+=0.07$ with either medium or low spatial input data.  
For the $\it inbetweening$ reconstruction, three time steps are considered; $\Delta T^+=12.6$ (medium), 25.2 (wide), and 126 (super-wide time step) in viscous time units, where $\Delta T^+$ is the time step between first and last snapshots.  
These $\Delta T^+$ correspond to temporal two-point correlation coefficients at $y^+=11.8$ of ${\cal R}=R^+_{uu}(t^+)/R^+_{uu}(0)\approx 0.50$ (medium), 0.25 (wide), and 0.05 (super-wide time step), {where 
$R^+_{uu}(t^+; y^+) = \overline{u^{\prime +}(x^+, y^+, z^+, \tau^+) u^{\prime +} (x^+, y^+, z^+, \tau^+ - t^+)}^{x^+,z^+,\tau^+}$, 
$\overline{\;\cdot\;}^{x^+,z^+,\tau^+}$ denotes the time-ensemble average over the $x-z$ plane, and $u^\prime$ is the fluctuation of streamwise velocity} \citep{FNKF2019}.  

	\begin{figure}
		\begin{center}
			\includegraphics[width=0.8\textwidth]{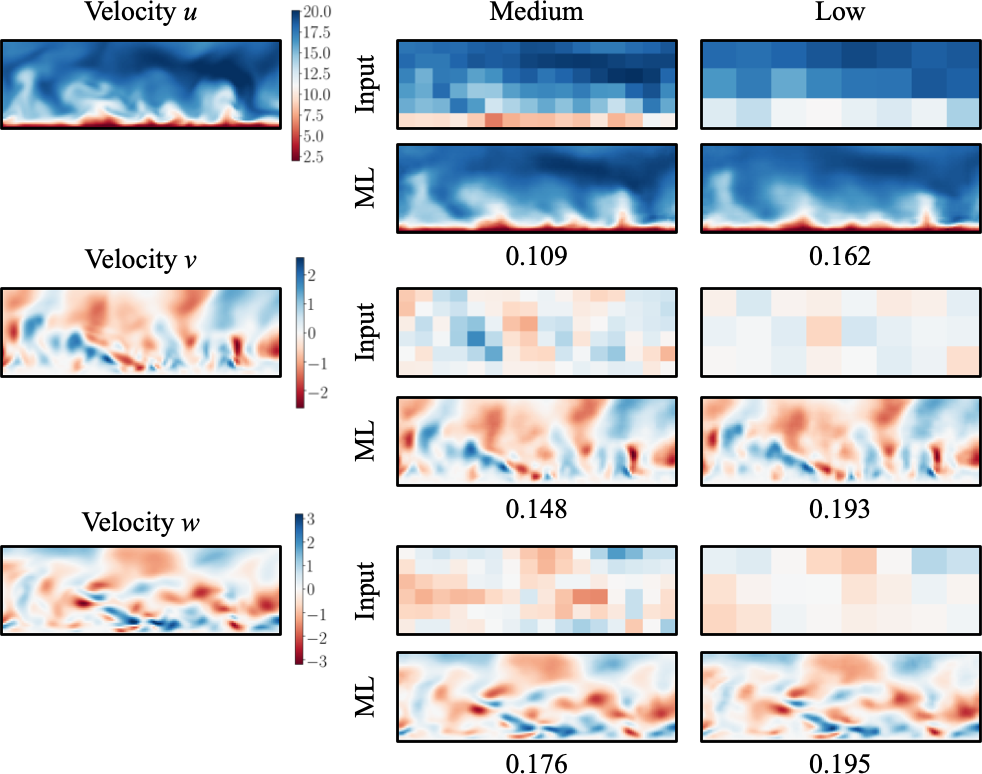}
			\caption{Velocity contours at a $y-z$ section ($x^+=1127$) of the reference DNS data, coarse input data, and the recovered flow field through spatial super resolution analysis.  The values listed below the contours are the $L_2$ error norm $\epsilon$.}
			\label{fig4}
		\end{center}
	\end{figure}
	\begin{figure}
		\begin{center}
			\includegraphics[width=0.9\textwidth]{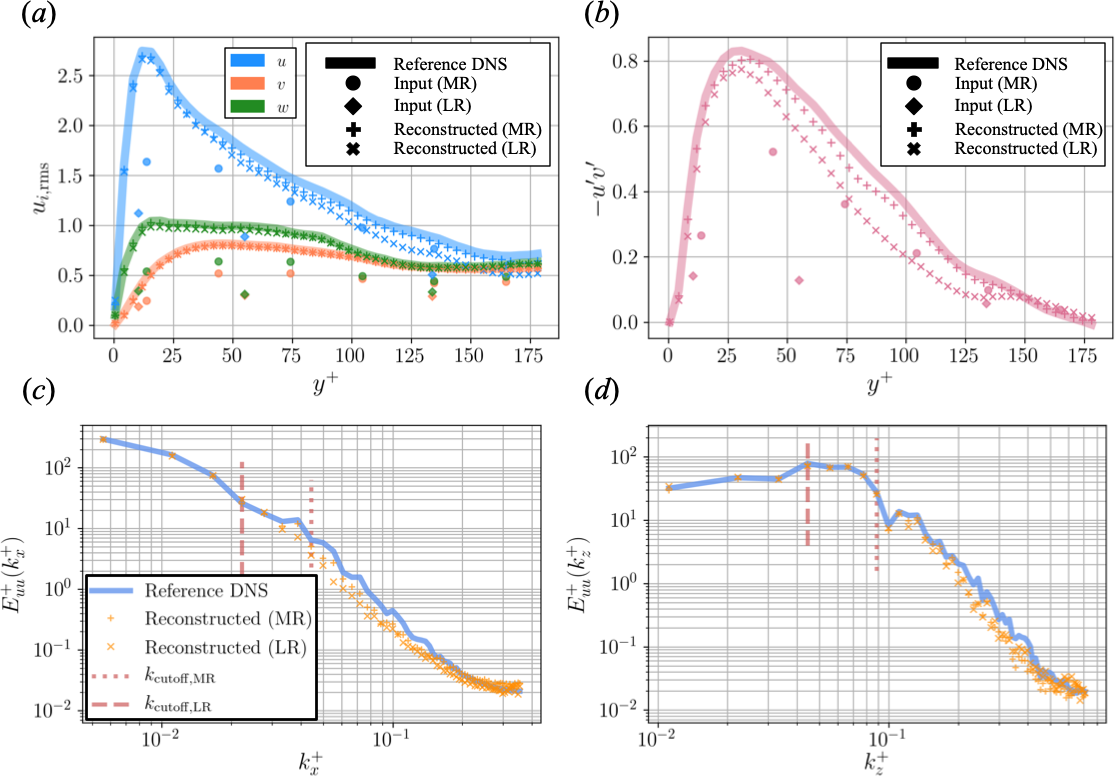}
			\caption{Turbulence statistics of the reference DNS data, input coarse data, and recovered flow fields by spatial super resolution analysis.  $(a)$ Root mean square of velocity fluctuation $u_{i,{\rm rms}}$, $(b)$ Reynolds stress $-u^\prime v^\prime$, $(c)$ streamwise energy spectrum $E^+_{uu}(k_x^+)$, and $(d)$ spanwise energy spectrum $E^+_{uu}(k_z^+)$.}
			\label{fig5}
		\end{center}
	\end{figure}

Let us summarize the reconstructed flow field visualized by the $Q$-criteria isosurface based on $n_{{\rm snapshots},x}=100$ in figure \ref{fig3}.  
The machine learning models are able to reconstruct the flow field from extremely coarse input data, despite the input data showing almost no vortex-core structures in the streamwise direction as shown in figure \ref{fig3}$(a)$.  
We also present the velocity contours at a $y-z$ section ($x^+=1127$) in figure \ref{fig4}.
We report the $L_2$ error norms normalized by the fluctuation component $\epsilon = ||{u}_{i,\rm DNS}-{u}_{i,\rm ML}||_2/||{u}^\prime_{i,\rm DNS}||_2$ in this example to remove the influence on magnitude of each velocity attribute in the present turbulent channel flow.
With both coarse input data, the reconstructed flow fields are in reasonable agreement with the reference DNS data in terms of the contour plots and the $L_2$ error norms listed below the reconstructed flow field.
We also assess the turbulence statistics as summarized in figure \ref{fig5}.  
Noteworthy here are the trends in the wall-normal direction that are well captured by the machine learning model from as little as 6 (medium) or 3 (low resolution) grid points, as shown in figures \ref{fig5}$(a)$ and $(b)$.  
Regarding the kinetic energy spectrum at $y^+=11.8$, the maximum wavenumber $k_{\rm max}$ in the streamwise and spanwise directions can also be recovered from the extremely coarse input data as presented in figures \ref{fig5}$(c)$ and $(d)$.  
The high wavenumber components for both cases {show differences} due to the fact that the dissipation range has no strong correlation with the energy-containing region.
{The aforementioned observation for the kinetic energy spectrum is distinct from that of two-dimensional turbulence.
The under or overestimation of the kinetic turbulent energy spectrum here is likely caused by a combination of several reasons, e.g., underestimation of $u$ because of the $L_2$ regression and relationship between a squared velocity and energy such that $\overline{u^2} = \int E^+_{uu}(k^+) dk^+$.
Whether machine-learning based approaches consistently yield underestimated or overestimated kinetic energy spectra likely depends on the flow of interest.  
While the overall method aims to minimize the loss function in the derivation, there are no constraints imposed for optimizing the energy spectra in the current approach. 
}

	\begin{figure}
		\begin{center}
			\includegraphics[width=1.0\textwidth]{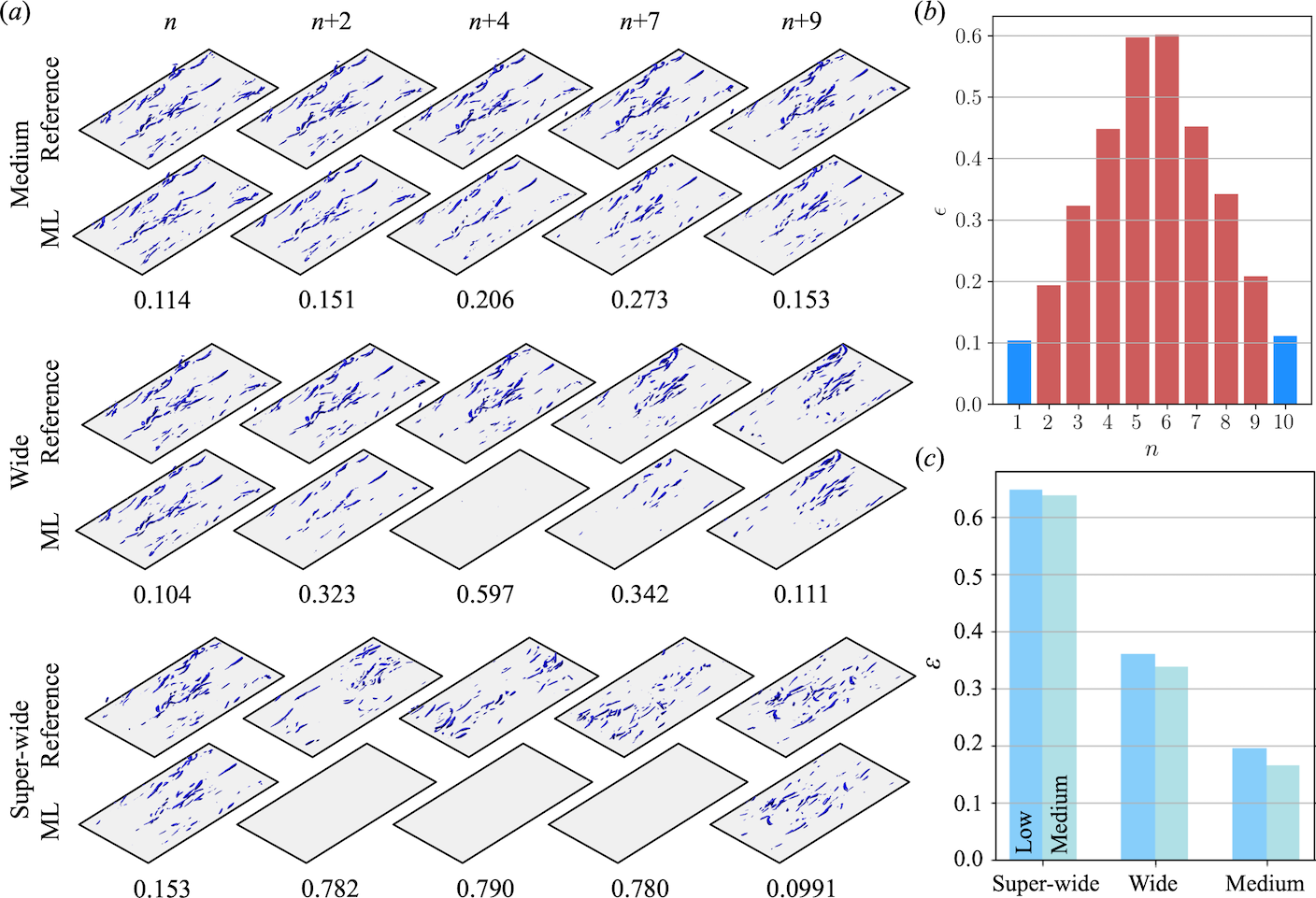}
			\caption{Spatio-temporal super resolution reconstruction of turbulent channel flow over three-dimensional domain.  $(a)$ The $Q$ isosurfaces ($Q^+=0.07$) of the reference DNS and super-resolved flow field with medium-, wide-, and super-wide time step.  The medium resolution data in space are used as the input for spatial super resolution reconstruction.  $(b)$ The $L_2$ error norm of inbetweening for spatial medium resolution input with wide time step.  $(c)$ Summary of the time-ensemble $L_2$ error norms for all combinations of coarse input data in space and time.}
			\label{fig6}
		\end{center}
	\end{figure}

Next, let us combine the spatial super resolution reconstruction with inbetweening to obtain the spatio-temporal high-resolution data ${\bm q}(x_{\rm HR},t_{\rm HR})$, as summarized in figure \ref{fig6}.  
We only show in figure \ref{fig6}$(a)$ the results for the medium spatial resolution input.
With the medium time step, the reconstructed flow fields show reasonable agreement with the reference DNS data in terms of both the $Q$ isosurface and $L_2$ error norm listed below the isosurface plots. 
In contrast, the flow fields cannot be reconstructed with wide- and super-wide time steps due to the lack of temporal correlation, as summarized in figure \ref{fig6_n}.
Although the vortex core can be somewhat captured with the wide time step at $n+2$ and $n+7$, the reconstructed flow fields are essentially smoothed since the machine-learned models for inbetweening are given only the information with low correlation at the first and last frames obtained from spatial super resolution reconstruction, as shown in figure \ref{fig6}$(b)$.
The time-ensemble $L_2$ error norm $\overline{\epsilon}$ with all combinations of coarse input data in space and time are summarized in figure \ref{fig6}$(c)$.  
It can be seen that the results with the machine-learned models are more sensitive to the temporal resolution than the spatial resolution level.  
This observation agrees with the previous example in section 3.1.

	\begin{figure}
		\begin{center}
			\includegraphics[width=\textwidth]{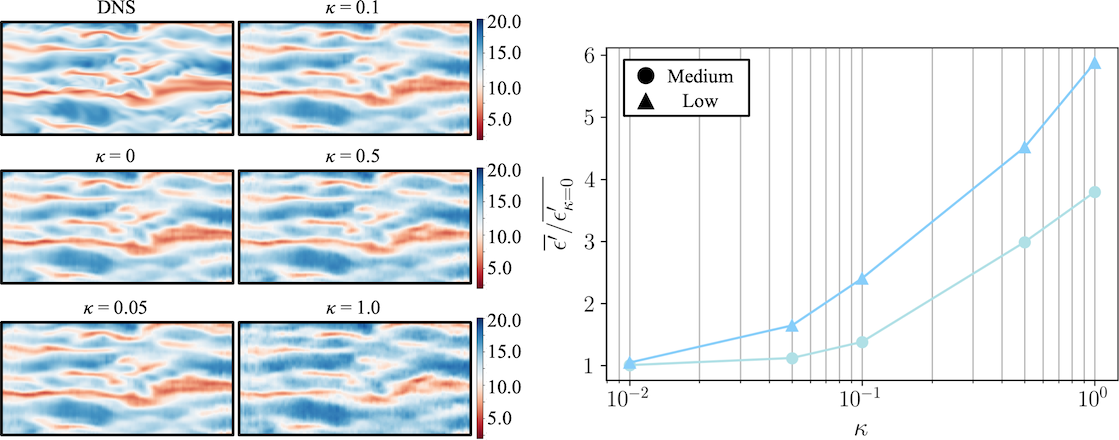}
			\caption{Robustness of the machine learning model for noisy input with medium time step models. The $x-z$ sectional streamwise velocity contours are chosen from $y^+=19.4$, with medium spatial coarse input model. {The contour plots visualize the intermediate snapshots at $t=(n+5)\Delta t$ for each noise level.}}
			\label{fig8}
		\end{center}
	\end{figure}

Let us demonstrate the robustness of the composite model against noisy input data for spatio-temporal super resolution analysis in figure \ref{fig8}.  
For this example, we use the medium spatially coarse input data with the medium time step.
Here, the $L_2$ error norm for noisy input is defined as $\epsilon^\prime_{\rm noise} = ||{\bm q}_{\rm HR}-{\cal F}({\bm q}_{\rm LR}+\kappa {\bm n})||_2/||{\bm q}^{\prime}_{\rm HR}||_2$, where $\bm n$ is the Gaussian noise {for which the mean of the distribution is the value on each grid point and the standard deviation scale is 1}, $\kappa$ is the magnitude of noisy input, and ${\bm q}^{\prime}_{\rm HR}$ is the fluctuation component of the reference velocity.
The reported values on the right side of figure \ref{fig8} are the ensemble-averaged $L_2$ error ratio against the original error without noisy input, $\overline{\epsilon^\prime}/\overline{\epsilon^\prime_{\kappa=0}}$.
As shown, the error increases with the magnitude of noise $\kappa$ for both coarse input levels.  
The $x-z$ sectional streamwise velocity contours from intermediate output of inbetweening at $t=(n+5)\Delta t$ are shown in the left side of figure \ref{fig8}. 
The model exhibits reasonable robustness for the considered noise levels, especially for reconstructing large-scale structures.

	\begin{figure}
		\begin{center}
			\includegraphics[width=0.50\textwidth]{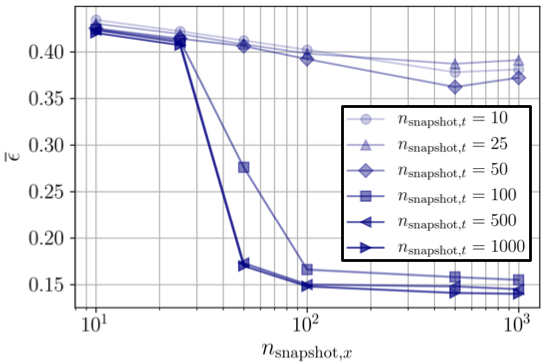}
			\caption{Influence of the number of the training snapshots for spatial super resolution reconstruction $n_{{\rm snapshot,}x}$ on the ensemble $L_2$ error norm $\overline{\epsilon}$.}
			\label{fig9}
		\end{center}
	\end{figure}

In the above discussions, we used 100 snapshots for both spatial and temporal super resolution analyses with three-dimensional turbulent flow.  
Here, let us discuss the dependence of the results on the number of snapshots for the spatial super resolution analysis $n_{{\rm snapshot,}x}$ and inbetweening $n_{{\rm snapshot,}t}$.
The ensemble $L_2$ error norm $\overline{\epsilon}$ with various number of snapshots is presented in figure \ref{fig9}.  
We summarize in this figure the effect from the number of training snapshots on the spatial and temporal reconstructions.
Up to $n_{{\rm snapshot},t}=50$, the $L_2$ errors are approximately 0.4 and the reconstructed fields cannot detect vortical structures from the $Q$-value visualizations, even if $n_{{\rm snapshot,}x}$ is increased.  
For cases with $n_{{\rm snapshot,}t}\geq 100$ and $n_{{\rm snapshot,}x}\geq 100$, the errors drastically decrease.  
For this particular example with the turbulent channel flow, one hundred training data sets is the minimum requirement for recovering the flow field for both in space and time.  
These findings suggest that data sets consisting with as few as 100 snapshots with the appropriate spatial and temporal resolutions hold sufficient physical characteristics for reconstructing the turbulent channel flow at this Reynolds number.

	\begin{figure}
		\begin{center}
			\includegraphics[width=0.55\textwidth]{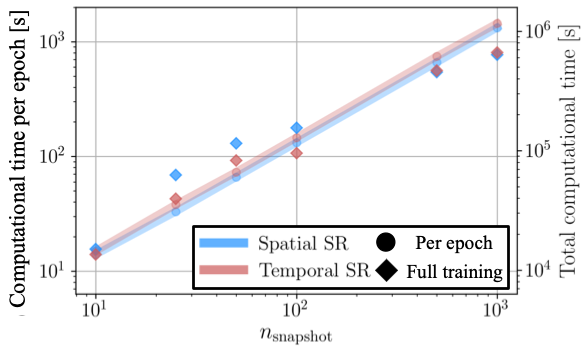}
			\caption{{
			Influence of the number of snapshots on the computational time per epoch (left) and total computational time (right).
			}}
			\label{fig16_20200423}
		\end{center}
	\end{figure}

Next, we assess the computational costs with increasing number of training data for the NVIDIA Tesla V100 graphics processing unit (GPU), as shown in figure \ref{fig16_20200423}.  
The computational time per an iteration ({epoch}) linearly increases with the number of snapshots for both spatial and temporal super resolution analyses.  
Plainly speaking, complete training process takes approximately 3 days with $\{n_{{\rm snapshot},x},n_{{\rm snapshot},t}\}=\{100,100\}$ and 15 days with $\{n_{{\rm snapshot},x},n_{{\rm snapshot},t}\}=\{1000,1000\}$.  
The computational costs for the full iterations can deviate slightly from the linear trend since the error convergence is influenced within the machine learning models due to early stopping.

Let us also discuss the challenges associated with the spatio-temporal machine learning based super resolution reconstruction.
As discussed above, supervised machine learning model is trained to minimize a chosen loss function through an iterative training process.
In other words, the machine learning models aim to solely minimize the given loss function, which is different from actually learning the physics.
We here discuss the dependence of the error in the real space and wave space.

	\begin{figure}
		\begin{center}
			\includegraphics[width=0.90\textwidth]{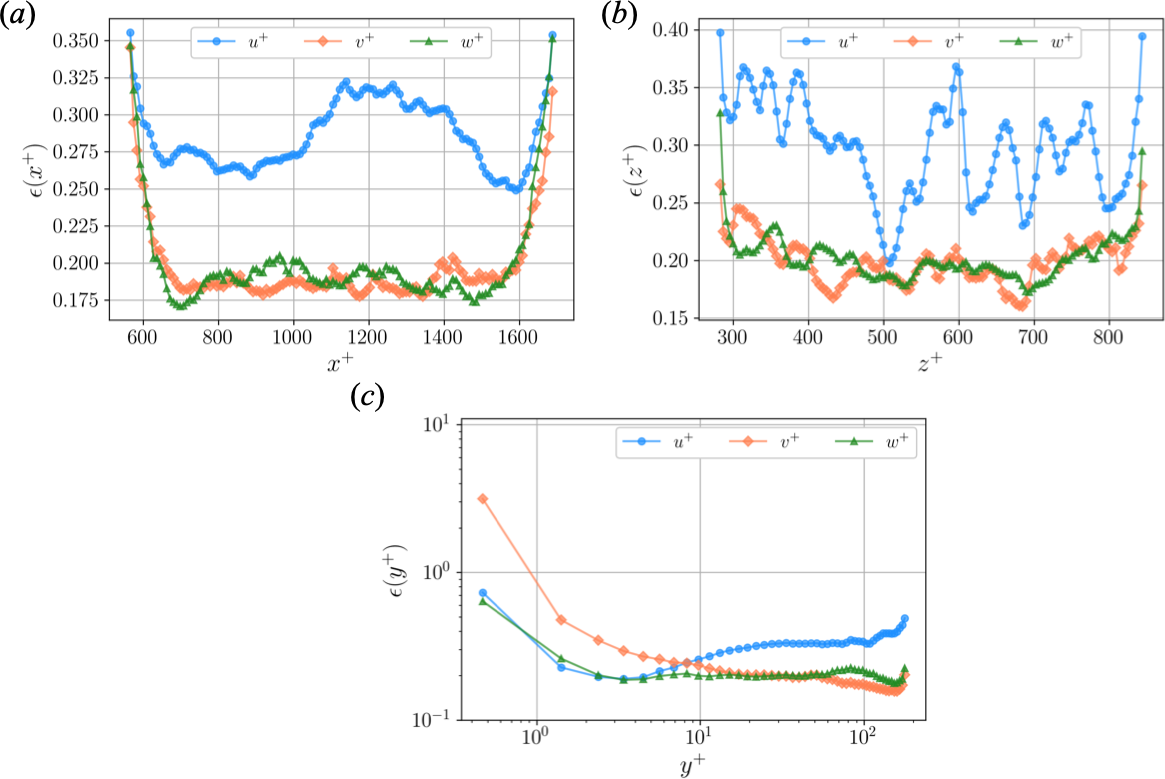}
			\caption{Dependence of $L_2$ error norm $\epsilon$ on location of turbulent channel flow in the $(a)$ streamwise ($x$), $(b)$ spanwise ($z$), and $(c)$ wall-normal ($y$) directions. For clarity, log-scale is used for $(c)$.}
			\label{fig18}
		\end{center}
	\end{figure}

The $L_2$ error distribution over each direction of turbulent channel flow is summarized in figure \ref{fig18}.
The case with combination of medium spatial input ($16\times 6 \times 16$ grids) with medium time step ($\Delta T^+=12.6$) is presented.
As shown in figure \ref{fig18}, the errors at the edge of the domain in all directions are large.
This is likely due to the difficulty in predicting the temporal evolution over boundaries and padding operation of convolutional neural networks.
Noteworthy here is the error trend in the wall-normal direction in figure \ref{fig18}$(c)$.
The errors for all attributes are high near the wall.
One of possible reasons is the low probability of velocity attributes near wall region{, i.e., high flatness factor \citep{KMM1987}}.
Since the present machine learning models are trained with $L_2$ minimization as mentioned above, it is tougher to predict those region than high probability for fluctuations.

	\begin{figure}
		\begin{center}
			\includegraphics[width=1.00\textwidth]{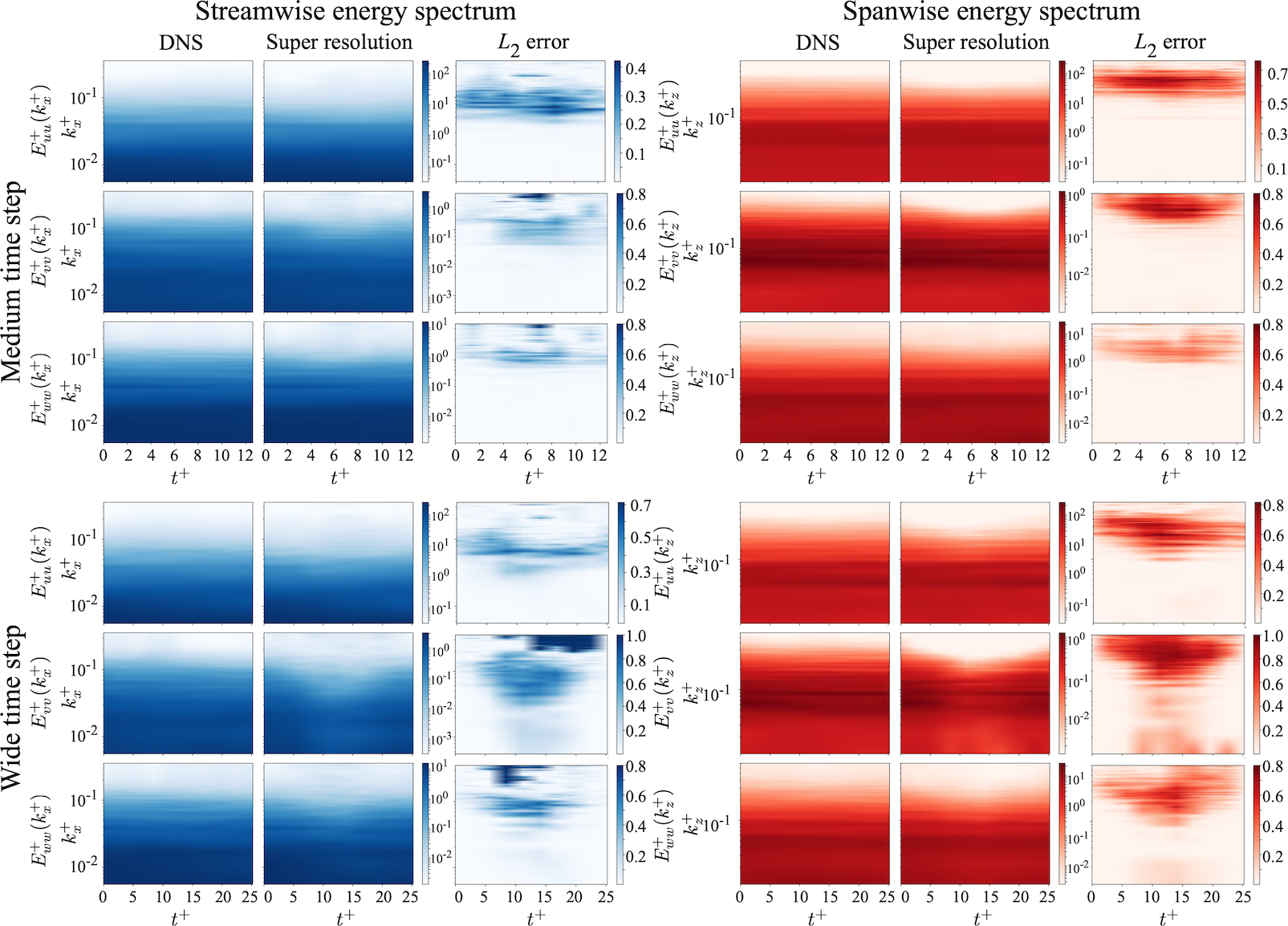}
			\caption{Streamwise and spanwise kinetic energy spectrum of spatio-temporal super resolution analysis. Top results are based on medium time step, and bottom results are based on wide time step. Left side presents the streamwise energy spectrum, and right side shows the spanwise energy spectrum.}
			\label{fig20}
		\end{center}
	\end{figure}

We further examine how well the machine learning model performs over the wavenumber space.
The kinetic energy spectra at $y^+=11.8$ in the streamwise and spanwise directions of spatio-temporal super resolution reconstruction are shown in figure \ref{fig20}.
For the input data, spatial medium resolution ($16\times 6\times 16$ grids) with medium- ($\Delta T^+=12.6$) and wide time steps ($\Delta T^+=25.2$) are considered.
The $L_2$ error here is defined as $\epsilon_{E^+_{uu}}(t^+)=||E^+_{uu}(t^+)_{\rm DNS}-E^+_{uu}(t^+)_{\rm ML}||_2/||E^+_{uu}(t^+)_{\rm DNS}||_2$.
With the medium time step, the error over the high-wavenumber space are higher than that over the low-wavenumber space.
This observation agrees with the machine learning models being able to recover the low-wavenumber space from grossly coarse data seen in figure \ref{fig5}.
With the use of wide time step, we can infer the influence of temporal coarseness, as discussed above.
The $L_2$ error distributions of kinetic energy spectrum show high-error concentration on high-wavenumber portion at intermediate output in time, as shown in the bottom part of figure \ref{fig20}.
The machine learning models capture low-wavenumber components preferentially to minimize the reconstruction error.

\section{Conclusion{s}}

We developed supervised machine learning method{s} for spatio-temporal super resolution analysis to reconstruct high-resolution flow data from grossly under-resolved input data {both in space and time}.  
First, a two-dimensional cylinder wake was considered as a demonstration.  
The machine-learned model was able to recover the data in space and reconstruct the temporal evolution from only {the} first and last frames.

As the first turbulent flow example, {a} two-dimensional decaying homogeneous isotropic turbulence was considered{.}
In this example, {we considered} two spatial {resolutions} based on our previous {work \citep{FFT2019b}} and two different time steps to examine the capability of the proposed model.  
The reconstructed flow fields were in reasonable agreement with the reference data in terms of the $L_2$ error norm, the kinetic energy spectrum, and the probability density function of vorticity field.  
We also found that the machine-learned models were affected substantially {by} the temporal range of training data.

We further examined the capability of the proposed method using {a} turbulent channel flow over three-dimensional domain at $Re_\tau=180$.  
The machine learning based spatio-temporal super resolution analysis showed its great capability to reconstruct the flow field from grossly {coarse input} data in space and time {when an} appropriate time step {size between} {the} first and the last frames {is used}.  
The proposed method, however, was unable to recover the turbulent flow fields in time {when} the temporal two-point correlation coefficient {was} ${\cal R}^+\leq 0.25$.
It was also seen that {the machine learning models tend to preferentially} extract the feature{s in the} low-wavenumber space so as to minimize a loss function efficiently. 
For improving the accuracy of the spatio-temporal super resolution analysis, we likely need to prepare a well-designed architecture which can take physics into {account in} its structure, {e.g.,} loss function {\citep{LY2019,raissi2020hidden,MFRFT2020}} and choice for input and output attributes, i.e., feature engineering.
{ In addition, care should also be taken for the proper choice of training data set which highly relates to the remaining problem --- the distinction of interpolation
and extrapolation for training data is still vague \citep{taira2019revealing}.}
Such efforts will be undertaken in future work. 

The robustness of the present model for noisy input and dependence on the amount of the training snapshots were also investigated.  
The proposed model showed reasonable capability for up to $10 \%$ noisy input in terms of both qualitative and quantitative assessments.  
We found that the flow field can be reconstructed by the machine learning-based methods with as little as 100 training data sets for {both the} spatial and temporal models.

We foresee a range of applications for the spatio-temporal super resolution analysis in fluid dynamics.  
For example, we may be able to leverage the current technique for large-eddy simulations as an augmentation tool.  
We may also be able to consider super resolution as a data compression tool to store big data{, which will also makes it easier to share or exchange spatio-temporal data of complex flows among researchers.}
In fact, in the present paper, we can recover the three-dimensional turbulent channel flow field comprised of 7.9 million spatio-temporal elements from low-resolution data which has only 3072 spatio-temporal elements.  This translates to a significant compression of $0.04\%$.

{
The ability to reconstruct turbulent flow fields from a small number of data points in space and time has the potential for novel data compression techniques using machine learning models.  
However, the success of flow field reconstruction hinges on developing an appropriate neural network construct that accommodates the complex nonlinear physics of turbulence. 
When successful, the neural networks are able to condense the flow field to its skeleton giving hope for autoencoder based approaches to identify the suitable coordinates to represent the primary axes of turbulent flow data in a nonlinear sense {\citep{FNF2020}}.
}

\section*{Acknowledgements}

Kai Fukami and Koji Fukagata thank the support from the Japan Society for the Promotion of Science (KAKENHI grant number: 18H03758).  
Kunihiko Taira acknowledges the generous support from the US Army Research Office (grant number: W911NF-17-1-0118) and the US Air Force Office of Scientific Research (grant number: FA9550-16-1-0650).  
Kai Fukami also thanks Takaaki Murata, Masaki Morimoto, Taichi Nakamura (Keio Univ.) and Kazuto Hasegawa (Keio Univ., Polimi) for insightful discussions.

\section*{Appendix}
	\begin{figure}
		\begin{center}
			\includegraphics[width=0.80\textwidth]{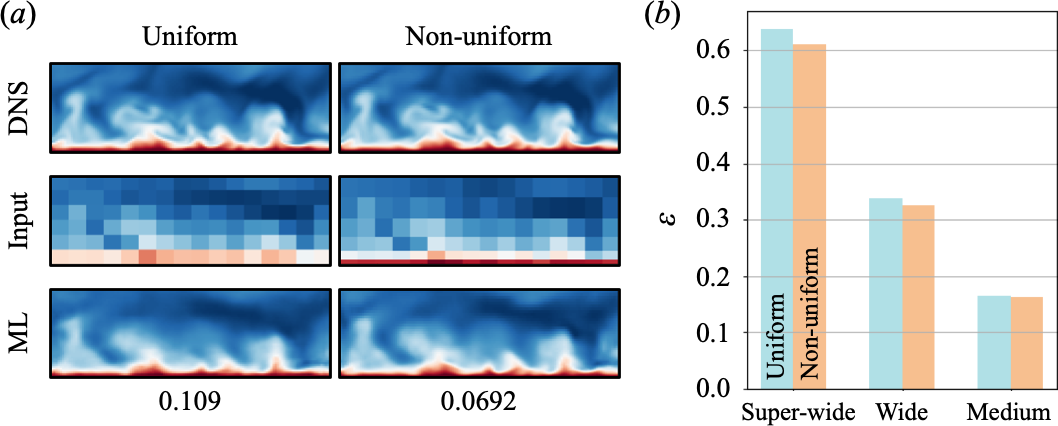}
			\caption{$(a)$ Dependence on the grid style of streamwise velocity contours $u$ at a $y-z$ section ($x^+=1127$) for spatial super resolution reconstruction.  Listed values are $L_2$ error norm. $(b)$ Time-ensemble $L_2$ error norm of spatio-temporal super resolution reconstruction with uniform and non-uniform grid data in wall-normal direction. }
			\label{fig10}
		\end{center}
	\end{figure}

In this appendix, we assess the influence of spatial discretizations in the wall-normal direction on the results from filter operation of the convolutional neural network.
These operations are generally performed on uniform resolution image data.  
Note that we use the interpolated flow fields on a uniform grid generated from the non-uniform grid data in $y$ direction, as input and output attributes for the discussions in the main text.

Here, we do not interpolate but instead use the data on a non-uniform grid.
We present the streamwise velocity contours at a $y-z$ section of spatial super resolution analysis with medium coarseness input data for both grid types in figure \ref{fig10}$(a)$.  
Although we observe some visual differences in the input data as shown in figure \ref{fig10}$(a)$, significant differences are not observed in terms of the streamwise velocity contours and the $L_2$ error norm.  
We also compare the $L_2$ error norm of the spatio-temporal super resolution analysis with both grid types in figure \ref{fig10}$(b)$.  
For all cases, the errors using the uniform grid data is slightly larger than those from the non-uniform grid data.
This is due to the original non-uniform data holding more information in the near-wall region compared to the uniform grid data.  
However, the filter operation of the convolutional neural network is not sensitive to the choice of the spatial discretization at least for this problem.  

\section*{Declaration of interest}

The authors report no conflict of interest.

\bibliographystyle{jfm}
\bibliography{jfm}

\end{document}